\documentstyle[prl,aps,multicol]{revtex}
\begin{document}
\title{\bf Hidden Kondo Effect in a Correlated Electron Chain}
\vspace{1.0em}
\author{A. A. Zvyagin$^{a,b}$ and H. Johannesson$^c$}
\address{$^a$Institut f\"ur Theoretische Physik, Universit\"at zu 
K\"oln, 
Z\"ulpicher Str. 77, 50937 K\"oln, Germany}
\address{$^b$B. I. Verkin Institute for Low Temperature Physics and 
Engineering  \\ of the National Academy of Sciences of Ukraine, 47 
Lenin Ave., Kharkov 310164, Ukraine}
\address{$^c$Institute of Theoretical Physics, Chalmers University of 
Technology and G\"oteborg University, S-412 96 G\"oteborg, Sweden}
\maketitle
\begin{abstract}
We develop a general {\em Bethe Ansatz} formalism for diagonalizing 
an integrable model of a magnetic impurity of arbitrary spin coupled 
ferro- or antiferromagnetically to a chain of interacting electrons. 
The method is applied to an open chain, with 
the exact solution revealing  
a ``hidden'' 
Kondo effect driven by forward electron 
scattering off the impurity. We argue that the so-called ``operator 
reflection matrices'' proposed in recent {\em Bethe Ansatz} studies 
of related models emulate only forward electron-impurity scattering 
which may explain the absence of complete Kondo 
screening for certain values of the impurity-electron coupling in 
these models. \end{abstract}
\centerline{PACS numbers 05.30.-d, 71.27.+a, 75.20.Hr}

\begin{multicols}{2}
\narrowtext

%\newpage
The study of magnetic impurities in one-dimensional (1D) strongly 
correlated electron or spin systems has attracted great interest in 
the last few years. The availability of nonperturbative methods in one dimension 
has allowed 
for a detailed picture of the relevant physics, revealing some rather 
unexpected features, such as the complete screening of an impurity 
spin for a {\em ferromagnetic} Kondo exchange ~\cite{FNFJ}. Future 
possible experiments on magnetic impurities implanted in quantum 
wires or carbon nanotubes, as well as analogies with related 
phenomena (X-ray boundary effects, metal point-contact 
spectroscopies, etc.), provide additional impetus for studying this 
problem. 

The {\em Bethe Ansatz} (BA) has played a particularly important role 
in the study of magnetic impurities. As is well-known, the method has 
successfully been employed for the exact treatment of a Kondo 
impurity in a free electron host as well as for mixed-valence 
impurities (with hybridized impurity and host wave functions) 
~\cite{obz}. The method has also been used to study magnetic 
impurities in spin chains ~\cite{AJ}, and more recently in {\em 
correlated} electron hosts ~\cite{mag}. In most of this work {\em 
periodic} boundary conditions (PC) were imposed on the electron (or 
spin) host. However, there exists an alternative approach, also 
exact, where {\em open} boundary conditions (OC) are implemented 
within a BA framework \cite{Sal}: A boundary potential at the edge of 
the system here plays the role of impurity scatterer. In a recent 
series of very interesting papers, Wang and collaborators 
~\cite{Wang} proposed several new BA solutions for magnetic 
impurities in correlated hosts with OC. In their approach the 
magnetic impurity is attached to the edge of the chain together with 
an auxiliary boundary potential that preserves integrability. The 
effect of the resulting composite edge on the electrons is coded in a 
``reflection matrix'', interpreted in ~\cite{Wang} as simulating 
back-scattering (BS) of electrons off an ordinary (nonintegrable) 
Kondo impurity in a correlated electron system. However, the absence 
of complete Kondo screening - as predicted by Furusaki and Nagaosa 
~\cite{FNFJ} - for certain values of the ferromagnetic exchange 
coupling in ~\cite{Wang} raises some concern about this 
interpretation.    

In this Letter we revisit the problem via an alternative route, 
exploiting the Quantum Inverse Scattering Method (QISM) \cite{KBI} to 
study the algebraic structure of this class of models. This allows us 
to explicitly exhibit the form of the reflection matrix used in 
~\cite{Wang} and show that it contains only forward electron 
scattering (FS) off the magnetic impurity, with the back-scattering 
against the (infinite wall) free edge potential playing no essential 
role for the physics of the impurity. This is different from a 
nonintegrable Kondo impurity in a correlated host, where the {\em 
dynamic} back-scattering against the impurity crucially influences the 
properties of the system ~\cite{FNFJ,LTFJ}. By a more general 
construction, valid for an integrable impurity spin of {\em 
arbitrary} magnitude coupled ferro- {\em or} antiferromagnetically to 
an electron host, we show that a Kondo effect is still operative also 
when the impurity appears to be unscreened: The associated ``hidden'' 
Kondo screening becomes manifest only in the presence of an external 
magnetic field or at nonzero temperatures. Most importantly, our 
analysis shows that forward electron-impurity scattering, without the 
assistance of backward scattering off a free edge potential, can 
drive Kondo screening in a correlated electron host. For 
transparency, we focus on the supersymmetric $t\!-\!J$ model of 1D 
correlated electrons with a spin-$S$ impurity. However, the principal 
results of our analysis hold for {\em any} integrable electron model 
with gapless low-lying excitations, and do not depend on the specific 
form of the host Hamiltonian. 

The key object in the QISM ~\cite{KBI} is the two-particle scattering 
matrix, $X_{a,m}(u)$, where $u$ is a spectral parameter, $a$ labels a 
subspace $V_a$ of an auxiliary particle and $m$ labels the Hilbert 
space $V_m$ of a particle at a site $m$ on a 1D lattice ~\cite{KBI}. 
The necessary and sufficient condition for integrability is that the 
scattering matrix satisfies the Yang-Baxter equation (YBE) 
$X_{ab}(u-v)X_{am}(u)X_{bm}(v) = X_{bm}(v)X_{am}(u)X_{ab}(u-v)$, 
implying that only {\em forward} scattering (FS) is allowed. To 
maintain integrability in the presence of an impurity, located at a 
site $n$ say, the impurity-host scattering matrix $R_{an}(u)$ also 
must satisfy a YBE: $X_{ab}(u-v)R_{an}(u)R_{bn}(v) = 
R_{bn}(v)R_{an}(u)X_{ab}(u-v)$. 

To set the stage, let us first look at the simplest case of an 
impurity in a spin-${1\over2}$ chain ~\cite{AJ}. The host as well as 
the impurity scattering matrices here belong to the SU(2)-symmetric 
rational solutions of the YBE: $X_{am}(u) = A(u)[u{\hat I} + ic{\hat 
P_{am}}]$ $(m=1,2,...,N)$ and $R_{an}(u) = A(u-i\alpha)[(u - i\alpha 
){\hat I} + ic{\hat P_{an}}]$, respectively. Here $c$ is a coupling 
constant (fixed by the YBE to be the same for host and impurity 
exchange), $|\alpha|$ measures the shift of the impurity level from 
the Kondo resonance ~\cite{mag}, $A(v)$ with $v=u, u-i\alpha$ are 
normalization constants, and $P_{aj}$ is a permutation operator on 
the corresponding spaces, with $V_j$ carrying a spin$-{1\over2}$ 
(spin-$S$) representation of $SU(2)$ for $j\!\! =\!\! m\!\! \neq \!\! 
n \ (j\!\! =\!\! n)$. Given $X_{am}(u)$ and $R_{an}(u)$ and imposing 
PC, QISM constructs the Hamiltonian of the system as a logarithmic 
derivative (w.r.t. to the spectral parameter) of the {\em 
transfer matrix} $\tau_{PC}(u)$ of the associated 2D statistical 
mechanics problem: $\tau_{PC}(u) \equiv Tr_a\prod_m^L 
X_{am}(u)R_{an}(u- i\alpha)$. It is important to note that the 
position of the impurity matrix in this product has no influence on 
the dynamics: The auxiliary particle simply scatters off all spins on 
the chain consecutively, including the impurity. For the OC case, one 
introduces additional {\em reflection} matrices, $K_a(u)$ 
~\cite{ChSkl} which describe the back-scattering off the open 
boundary. In contrast to the host or impurity scattering matrices, 
these are $c$-number matrices. They satisfy the reflection equation 
(RE) $X_{ab}(u-v)K_a(u)X_{ab}(u+v)K_b(v) = 
K_b(v)X_{ab}(u+v)K_a(u)X_{ab}(u- v)$, as required by integrability. 
Given the reflection matrices, the analog of the transfer matrix for 
the OC, $\tau_{OC}(u)$, is defined by $\tau_{OC}(u) = Tr_a 
K_a(u)T_a(u)K_a(u)T_a^{-1}(-u)$, where $T_a(u) = \prod_m^L 
X_{am}(u)R_{an}(u-i\alpha)$ is the PC monodromy matrix. The 
recently proposed ``operator reflection matrix'' for the spin model 
in ~\cite{Wang} has the simple structure $R(u)K(u)R^{-1}(-u)$ with 
$K(u)={\hat I}$, i.e. it is just the ordinary $c$-number reflection 
matrix $K(u)$ of a free boundary sandwiched between two FS impurity 
matrices ~\cite{FZ,Wang}. The auxiliary particle here scatters off 
the impurity, reflects at the free edge, and then scatters off the 
impurity once more, but moving in the opposite direction. 

The QISM for correlated PC ~\cite{EK} or OC ~\cite{GRE} electron 
chains with an impurity works similar to the scheme above, with one 
essential difference: electrons carry spin {\em and} charge, and, 
hence, two {\em nested} transfer matrices have to be introduced. The 
first-level transfer matrix describes the charge sector, while the 
second-level transfer matrix describes the spin sector. Because of 
the nesting, a magnetic impurity inserted into a correlated electron 
chain has to carry both spin {\em and} charge degrees of freedom in 
order to preserve integrability. Its spin part drives the Kondo 
effect while the charge part provides the mixed-valence behavior of 
an impurity ~\cite{two}.

Specializing to the supersymmetric $t\!-\!J$ model ~\cite{SUSY} with 
a magnetic impurity, its Hamiltonian can be decomposed as ${\cal H} = 
{\cal H}_{bulk} + {\cal H}_{imp} + {\cal H}_{bound}$. Here 
${\cal H}_{bulk} = K_{\alpha \beta}\sum_{n=1}^{L-1}(J_n^{\alpha} 
J_{n+1}^{\beta} + h.c.)$ defines the bulk Hamiltonian for a chain of 
length $L$, with $J_n^{\alpha}$ the generators in the defining 
representation of the supersymmetric algebra $sl(1|2)$, and 
$K_{\alpha \beta} \equiv Tr J^{\alpha}J^{\beta}$ ~\cite{EK}. The 
impurity Hamiltonian ${\cal H}_{imp}$ (with the impurity coupled to 
sites $n$ and $n+1$) has the form  
\begin{eqnarray}
&&{\cal H}_{imp} = c_0\bigl(H_{n,S} + H_{S,n+1} - (\alpha^2 + 
2S(S+1))H_{n,n+1}  \nonumber \\
&&- 2i\alpha[H_{S,n} + H_{S,n+1},H_{n,n+1}] + \{H_{n,S},H_{S,n+1}\} 
\bigr).
\label{Himp}
\end{eqnarray} 
The commutator-anticommutator structure in (\ref{Himp})  
is generic and applies to {\em any} impurity model (with $SU(2)$ or $sl(n|m)$ 
symmetry) constructed by QISM. It is here realized by taking $H_{n,S} 
= K_{\alpha\beta}(J_n^{\alpha}J_S^{\beta} + h.c.)$, where 
$J_S^{\alpha}$ are the generators for the spin-$S$ impurity, with 
$c_0 = f[(S+{1\over2})^2 - \alpha^2]^{-1}$ being an effective impurity-host 
coupling constant ($f=1$ for an exchange
impurity, while $f = \bigr(M \sigma| M + \sigma \bigl)\sqrt{2S+1}$ 
for a hybridization impurity, with $\bigr(M \sigma | M + \sigma 
\bigl)$ the Clebsch-Gordan coefficients ~\cite{mag}). The boundary 
Hamiltonian ${\cal H}_{bound}$ has a trivial structure for PC: ${\cal 
H}_{bound} = K_{\alpha \beta}J_1^{\alpha} J_{L}^{\beta} + h.c$. 
The OC boundary Hamiltonian, on the other hand, is obtained by making 
the replacement $(J^{\alpha})_{1,L} \to h_{1,L}$, where $h_{1,L}$ 
define the {\em boundary fields} at the edges at $m=1$ and $m=L$ 
\cite{one}. This procedure is directly applicable when the impurity 
is located in the bulk. However, a similar construction can be used 
also for an impurity {\em at the edge}: We now put $n=L$, and replace 
the operator at the ``phantom site'' with index $n+1$ in Eq. (\ref{Himp}) 
by the boundary field $h_{L}$. Note that by this procedure the three-particle
commutator and anticommutator terms in (\ref{Himp}) collapse 
to two-particle terms.

Inspection of Eq. (\ref{Himp}) shows that the parameter $\alpha$ 
determines the coupling between impurity and host. For imaginary 
$\alpha$, and for real $\alpha$ with $|\alpha| < S+{1\over2}$, we 
have an antiferromagnetic (AFM) coupling, while for real $\alpha$ 
with $|\alpha| > S+{1\over2}$ we get a ferromagnetic (FM) coupling. A 
real $\alpha$, however, corresponds to a non-Hermitian impurity 
Hamiltonian, making the ferromagnetic case unphysical unless one 
places the impurity at the edge with a {\em zero boundary field} 
~\cite{mag,FZ}. For this special choice, $h_L=0$, only the first term 
survives in Eq.~(\ref{Himp}): ${\cal H}_{imp} = c_0H_{L,S}$. Thus, 
the FS impurity is here connected to the host by a single link with 
coupling constant $c_0$, providing a simple and natural impurity 
Hamiltonian. Analogous to the spin-chain case, the ``reflection 
matrix'' including the impurity is obtained by sandwiching the 
ordinary {\em free edge} reflection matrix $K(u)=\hat{I}$ between two 
FS impurity matrices: $R_{aL}(u)\hat{I}R_{aL}(-u)$ with $R_{aL}(u)$ 
from ~\cite{mag}. This structure is general and holds also for the 
models considered in ~\cite{Wang} as is evident from inspection of 
the resulting BA equations. Its form implies that the back-scattering 
from the free edge, which is present in any open chain, decouples 
from the scattering governed by the FS impurity matrices. As a 
consequence, the position of the impurity on the chain is immaterial 
to the physics when the interaction is AFM with imaginary $\alpha$. 
On the other hand, as we have just seen, for real $\alpha$ (including 
FM interaction) a real energy spectrum requires the impurity to be 
attached to the edge {\em with zero boundary potential}.

Eigenfunctions and eigenvalues of the model are parameterized by sets 
of quantum numbers, partitioned into {\em charge rapidities}  
$\{u_j\}_{j=1}^N$ (with $N$ the number of electrons) and {\em spin 
rapidities}, $\{v_q\}_{q=1}^M$ ( with $M$ the number of ''down 
spins''). The rapidities are the solutions of the BA equations, which 
for the OC zero-boundary case take the form:    
\begin{eqnarray}
&&\prod_{\pm}e_{2S\pm \alpha}(v_p)\prod_{j=1}^N e_1(v_p \pm 
u_j) = \prod_{\pm}\prod_{q=1}^M e_2(v_p \pm v_q) 
\nonumber \\
&&e_1^{2L}(u_j) = 
\prod_{\pm}Y_{\pm}(u_j)\prod_{p=1}^M e_1(u_j \pm v_p) \ , 
\label{BAE}
\end{eqnarray} 
with $e_n(x) = (2x + in)/(2x - in)$. The functions containing 
$\alpha$ describe spin and charge degrees of freedom of the impurity 
($Y_{\pm}(x) = e_{2S+1\pm \alpha}(x)$ for a hybridization impurity 
and $Y_{\pm}(x) = \sqrt{e_{2S+1\pm\alpha}(x)/e_{2\pm\alpha}(x)}$ for 
an exchange impurity ~\cite{mag}). These BA equations can be 
transformed into a form similar to the PC case by a change of 
variables: $u_j \to - u_j$, $j = - N,..., -1, 0$;  $v_p \to - v_p$, 
$p = -M,..., -1, 0$ which gives the OC energies $E = 
\sum_{j=1}^{2N+1}(u_j^2 + {1\over 4})^{-1}$. We also remove the 
roots corresponding to $u_j = v_p = 0$ (which label unphysical null 
states). It is important to stress that the states which are present 
for OC but not PC determine the BS singularities {\em independent} of 
the FS {\em impurity} terms. 

The groundstate of the supersymmetric $t\!-\!J$ model in an external 
field is obtained by filling up two Dirac seas for singlet Cooper-
like pairs and unbound electrons, respectively ~\cite{dif}. The 
structure of the singlet-paired groundstate for zero magnetic field 
$H=0$ conspires with the magnetic impurity to produce a nonzero {\em 
mixed} impurity valence $n$: For $H=0$ there are no unbound 
electrons, but scattering of Cooper pairs off the exchange 
(hybridization) impurity makes $n$ smoothly vary from zero for an 
empty band to $+1$ $(-1)$ for a half-filled band, a process common to 
both FM and AFM impurity-host coupling. By an analysis of the 
counting functions that define the number of BA states ~\cite{dVW} 
one can show that the impurity magnetization $M_{imp}$ in the 
limit of {\em zero magnetic field} can take either the value 
$M_{imp}=S-{1\over2}$ (as in the ordinary Kondo effect) or 
$M_{imp}=S$. In the latter case the Kondo screening is ``hidden'', 
and becomes manifest only for nonzero fields or temperatures. This 
effect, which is generic to this class of theories, is particularly 
transparent in the present model. 

To see how it comes about, let us first consider the case with AFM 
impurity-host coupling and imaginary $\alpha$. Here the impurity 
``traps'' a fraction of a Cooper pair which gets polarized by the 
field to produce an {\em effective} impurity spin $S_{eff} = 
S+{|n|\over2}$. However, the magnetic field also excites unbound 
electrons from the sea of Cooper pairs. For a sufficiently weak 
field, these unbound electrons partially screen the effective 
impurity spin, a process in complete analogy with the ordinary Kondo 
effect with the only difference that an {\em effective spin} $S_{eff} 
> S$ gets (partially) screened. As the field increases it eventually 
breaks up the impurity-screening cloud composite, leaving behind the 
effective (unscreened) spin $S_{eff}$. For $S>{1\over2}$ there is a 
crossover between low- and high-energy behaviors of the magnetic 
impurity: For low fields one has an asymptotically free underscreened 
spin $S$, while for high fields the asymptotically free spin is 
$S+{n\over2}$. We can also see the features of the hidden Kondo 
effect in the finite-temperature properties. For example, in the 
Kondo regime (with charge degrees of freedom suppressed) the 
effective spin is $S$ for low temperatures, $T\ll T_K$, and 
$S+{1\over2}$ for high temperatures, $T\gg T_K$, with Curie-like 
behavior and usual Kondo logarithmic corrections. The zero field 
residual entropy is given by ${\cal S} = \ln 2S$ for imaginary 
$\alpha$. The specific heat has a Shottky peak at $T \propto H$ and a 
Kondo resonance at $T \propto T_K$ for a weak magnetic field $H$, 
with the two peaks merging into one for large $H$. This behavior is 
typical for an ``underscreened'' magnetic impurity ~\cite{obz}. In 
contrast, {\em complete} Kondo screening is present for the case of 
an exchange impurity with $S={1\over2}$ or a hybridization impurity 
with $S=0$. The impurity susceptibility is proportional to $T_K^{-
1}$, with a specific heat linear in $T$ at low energies, and one thus 
recovers a standard Fermi-liquid scenario generic for AFM impurity 
models. 

We can illustrate the above, e.g. by explicitly calculating the 
impurity magnetization $M_{imp}$ at half-filling, using the BA 
equations (\ref{BAE}). In fact, we find a universal expression for 
$M_{imp}$, {\em valid for the AFM as well as the FM case}:    
\begin{equation}
M_{imp} =S_{eff} \bigl[1 \pm {1\over 2\ln (H/T_K)} - {\ln \ln 
(H/T_K)\over 4 \ln^2 (H/T_K)} + \dots \bigr] \ , 
\label{magn}
\end{equation}
where for imaginary $\alpha$ the Kondo energy scale is $T_K = H_0 
\exp (-\pi |\alpha|)$ with $H_0 = \sqrt{\pi^3/e}$, and where we have 
subtracted the contribution $M_{edge} = |2\ln (H/H_0)|^{-1} - \ln 
\ln|\sqrt{H/H_0}|/4\ln^2 (H/H_0) +...$ from the free edges. For low 
fields, and $S>{1\over2}$, $H \ll T_K$, $S_{eff} =S$ with the upper 
sign in (\ref{magn}) defining $M_{imp}$. On the other hand, for 
fields which are large on the Kondo scale but still much smaller than 
the spin saturation field, $T_K \ll H \ll 1$, $S_{eff}= S+{1\over2}$ 
with $M_{imp}$ defined by the lower sign in (\ref{magn}). For an 
$S=0$ hybridization impurity or an $S={1\over2}$ exchange impurity 
one obtains $S_{eff} \propto H/T_K$ for low magnetic fields, while 
$S_{eff} = {1\over2}$ for high fields \cite{mag}. 

Let us now consider the FM case, or more generally, the case of real $\alpha$,  
with $2|\alpha| = [2|\alpha|] + \{ 2|\alpha|\}$, where $[x]$ ($\{ x\}$) 
denotes the integer (fractional) part of $x$. Eq.~(\ref{magn}) still 
describes the impurity magnetization, with the fractional part 
determining the ``Kondo temperature'': $T_K = H_0 /\cos (\pi 
\{2|\alpha|\}/2)$. This is in contrast to the case with imaginary 
$\alpha$ which exhibits a usual exponential dependence of $T_K$ on 
the coupling constant. (Note that for $2|\alpha|$ an integer we have 
$T_K = H_0$.) Hence the crossover scale is here larger, $T_K > 1$, 
than the critical field that determines the transition to the 
spin-saturated, ferromagnetic phase of the host. It follows that for 
real $\alpha$ a {\em high-field region for a magnetic impurity is 
absent} ~\cite{nec}. Another feature special for real $\alpha$ is 
that incident and reflected particles effectively scatter off {\em 
different} impurity spins: $S \pm {[2|\alpha|]\over2}$ and $S+{1\pm 
[2|\alpha|]\over2}$ respectively. Thus, depending on the value of 
$\alpha$, the impurity exhibits different characteristics. If 
$[2|\alpha|] < 2S-1$, $[2|\alpha|]= \pm 2S$, or $[2|\alpha|] > 2S+1$ 
(FM domain), then $S_{eff}=S$ in Eq.~(\ref{magn}): the impurity spins 
``seen'' by incident and reflected waves are both underscreened. 
Thus, for the FM regime, the impurity spin is {\em always} 
underscreened, also for $S={1\over2}$. This suggests that the 
complete screening for FM Kondo coupling as proposed in ~\cite{FNFJ} 
crucially depends on the presence of back-scattering for this case.  
For the special values $[2|\alpha|]=2S\pm 1$, 
$S_{eff} =S\pm {1\over4}(1 - {H\over T_K})$ we obtain {\em both 
a remnant spin and terms linear in $H$}. Similar features also appear 
for the specific heat for these critical values: The Curie-like 
behavior of the remnant spin (the underscreened effective spin seen 
by incoming waves) is accompanied by a Fermi-liquid behavior (of the 
totally screened effective spin seen by reflected waves) 
~\cite{frac}. Negative effective spins signal the appearance of local 
levels (bound states of host excitations). These levels, which 
influence the value of the remnant impurity entropy, are generated by 
the FS magnetic impurity, and are insensitive to the edge potential. 
We point out that $\alpha$ determines the shift of the Kondo 
resonance, with imaginary $\alpha$ (AFM coupling) corresponding to a 
resonance with the band excitations of the host, while for real 
$\alpha$ (FM {\em or} AFM coupling) local impurity levels may 
decouple from the bands. 

To conclude, we have shown that forward electron-impurity scattering 
in a correlated host can drive a Kondo effect {\em without} the 
assistance of backward scattering from a free edge potential. This 
Kondo effect, which is present both for ferro- {\em and} 
antiferromagnetic impurity- electron coupling, is hidden for impurity 
spin $S>{1\over2}$ (as well as for $S={1\over2}$ when the coupling is 
ferromagnetic) and becomes manifest only in the presence of a 
magnetic field or at nonzero temperatures. We have argued that the so-
called ``operator reflection matrices'' proposed in recent {\em Bethe 
Ansatz} studies of related models ~\cite{Wang} emulate only forward 
scattering off a magnetic impurity. This may explain the observed 
absence of complete Kondo screening for certain values of the 
Kondo coupling in these models. 

The authors acknowledge support from Deutsche Forschungsgemeinschaft 
(A.A.Z.) and the Swedish Natural Science Research Council (H.J.).

\end{multicols}

\begin{references}
\bibitem{one} The boundary field couples to the $sl(1|2)$ generators 
and hence plays the role of a magnetic field in the spin sector. In 
addition, a nonzero $h_{1,L}$ acts in the charge sector to produce a 
nonzero phase between incoming and reflected electron waves. Here we 
use the case with the same boundary potentials for spin and charge 
sectors, ~\cite{GRE}.
\bibitem{dif} This is different from the $t\!-\!J$ model studied in 
~\cite{Wang}, where the two Dirac seas correspond to unbound 
electrons and spinons.
\bibitem{nec} As a consequence, a nonperturbative approach - as in 
the present Letter - is required since perturbation theory is known 
to break down at low energies in this class of problems [2].
\end{references}
\end{document}